\documentclass[aps, pra, twocolumn, 10pt, longbibliography,floatfix]{revtex4-2}
\usepackage{soul}
\usepackage{microtype}
\usepackage{amsmath}   
\usepackage{amssymb}
\usepackage{mathtools}
\usepackage{graphicx}
\usepackage{dcolumn}
\usepackage{bm}
\usepackage{amsmath}
\usepackage{graphicx} 
\usepackage{amsfonts}
\usepackage{subfigure}
\usepackage{graphicx}
\usepackage{array}
\usepackage{color}
\usepackage{amssymb}%

\usepackage{amsmath,mathrsfs}
\usepackage{placeins}

\newcommand{\rbisoa}[1]{\(^{87}\text{Rb}\)#1}
\newcommand{\rbisob}[1]{\(^{85}\text{Rb}\)#1}

\begin{document}

\title{Short-term stability of a microcell optical reference based on Rb atom two-photon transition at 778 nm}
    
\author{Martin Callejo ${}^1$}
\author{Andrei Mursa ${}^1$ }
\author{ R\'emy Vicarini ${}^1$}
\author{Emmanuel Klinger ${}^1$}
\author{ Quentin Tanguy ${}^1$}
\author{Jacques Millo ${}^1$}
\author{Nicolas Passilly ${}^1$}
\author{Rodolphe Boudot ${}^1$}

\affiliation{${}^1$ FEMTO-ST, CNRS, Universit\'e de Franche-Comt\'e, ENSMM, 26 chemin de l'Epitaphe 25030 Besan\c{c}on Cedex, France}

\affiliation{Corresponding author: rodolphe.boudot@femto-st.fr}

\date{\today}



\begin{abstract}
{\bfseries
We report on the development and short-term stability characterization of an optical frequency reference based on the spectroscopy of the rubidium two-photon transition at 778 nm in a microfabricated vapor cell. When compared against a 778 nm reference signal extracted from a frequency-doubled cavity-stabilized telecom laser, the short-term stability of the microcell frequency standard is 3.5 $\times$ 10$^{-13}$ $\tau^{-1/2}$ until 200~s, in good agreement with a phase noise level of $+$ 43 dBrad$^2$/Hz at 1~Hz offset frequency. The two main contributions to the short-term stability of the microcell reference are currently the photon shot noise and the intermodulation effect induced by the laser frequency noise. With still a relevant margin of progress, these results show the interest of this spectroscopic approach for the demonstration of high-stability miniaturized optical vapor cell clocks.
Such clocks are poised to be highly beneficial for applications in navigation, communications, and metrology.
}
\end{abstract}


\maketitle

\section{Introduction}\label{introduction}
 
About twenty years ago, the advent of microfabricated alkali vapor cells \cite{Kitching:APL:2002}, combined with low-power and high-bandwidth vertical-cavity surface emitting-lasers (VCSELs) \cite{Affolderbach:VCSELs}, enabled the demonstration of the first microfabricated atomic clock \cite{Knappe:APL:2004}. %
The physics package of this microwave clock, based on coherent population trapping (CPT) \cite{Arimondo:1996, Cyr:TIM:1993}, had a volume of 9.5 mm$^3$, consumed 75 mW, and demonstrated a fractional frequency stability of 2.5~$\times$~10$^{-10}$~$\tau^{-1/2}$ up to 30 seconds ($\tau$ being the integration time), and of 2 $\times$ 10$^{-9}$ at 10$^4$~s. 
%

Following this prototype, intense research efforts were undertaken to enhance the performances of these clocks, including the development of VCSELs tuned on the alkali D$_1$ line  \cite{Stahler:OL:2002, Serkland:SPIE:2007, AlSamaneh:APL:2012, Kroemer:AO:2016}, MEMS cell technologies with improved internal atmosphere \cite{Knappe:OL:2005, Douahi:EL:2007, Liew:APL:2007, Hasegawa:SA:2011, Karlen:OE:2017, Maurice:APL:2017, Vicarini:SA:2018, Bopp:JPP:2021, Maurice:MSNE:2022}, low noise and low power electronics \cite{Zhao:TIM:2014}, or integrated clock physics packages \cite{Lutwak:PTTI:2007, Haesler:CMAC:2017}.

These advances resulted in the emergence of commercially-available  chip-scale atomic clocks (CSACs) with a volume of 15-20 cm$^3$, a power consumption lower than 150 mW, and a timing error of a few microseconds at 1 day \cite{Kitching:APR:2018}. These clocks are now deployed for underwater sensing, navigation systems, or secure and jam-resistant communications. 

More recently, prototype microcell CPT clocks have showcased  frequency stability levels in the low 10$^{-12}$ range at 1~day \cite{Zhang:IEEEULPAC:2019, Yanagimachi:APL:2020, Carle:OE:2023} thanks to advanced light-shift compensation techniques \cite{Yanagimachi:APL:2020, MAH:APL:2022} and microcells built with low gas permeation glass windows \cite{Dellis:OL:2016, Carle:JAP:2023, Kozlova:PRA:2011}. 

Other approaches have also been explored for the advent of miniaturized microwave clocks with enhanced stability. A microcell clock based on the pulsed optically pumped (POP) technique \cite{Micalizio:Metrologia:2012}
was demonstrated in \cite{Batori:PRAp:2022} with a stability of 1.5~$\times$~10$^{-11}$ at 1~s and in the low 10$^{-12}$ range at 1~day. In \cite{Martinez:NC:2023}, a chip-scale beam microwave clock was reported with a stability of 1.2~$\times$~10$^{-9}$~$\tau^{-1/2}$, averaging down until 250~s.


A stimulating approach for demonstrating miniature clocks with significantly enhanced short-term stability consists of interrogating optical transitions in a microfabricated vapor cell, instead of microwave transitions. In this context, sub-Doppler spectroscopy techniques, based on the interaction of hot alkali atoms with two counter-propagating beams provided by a single laser source, are particularly attractive due to their extreme simplicity.

Saturated absorption spectroscopy (SAS) in a Rb microcell was reported in \cite{Hummon:O:2018} using a Distributed Bragg Resonator (DBR) laser, yielding a frequency stability below 10$^{-11}$ up to 10$^4$~s. A compact optical module, with a volume of 35~cm$^3$, a weight of 35~g, and a power consumption of 780~mW, was also demonstrated with SAS using a small-size glass-blown cell \cite{Strangfeld:JOSAB:2021}. Piloted by embedded FPGA electronics, this reference reached a fractional frequency stability of 1.7~$\times$~10$^{-12}$ at 1~s and 6~$\times$~10$^{-12}$ at 10$^5$~s \cite{Strangfeld:OE:2022}. Dual-frequency sub-Doppler spectroscopy (DFSDS) \cite{MAH:OL:2016, DB:PRA:2019}, that relies on a setup comparable to SAS but using a dual-frequency optical field, demonstrated, when applied to a Cs microcell with an external-cavity diode laser (ECDL), stability performances of 3~$\times$~10$^{-13}$ at 1~s and below 5~$\times$~10$^{-14}$ at 100~s \cite{Gusching:OL:2023}. Nevertheless, DFSDS requires the generation of a microwave-modulated laser, thereby introducing greater complexity to the optical setup architecture.



\begin{figure}[t!]
\centering
{\includegraphics[width=0.86\linewidth]{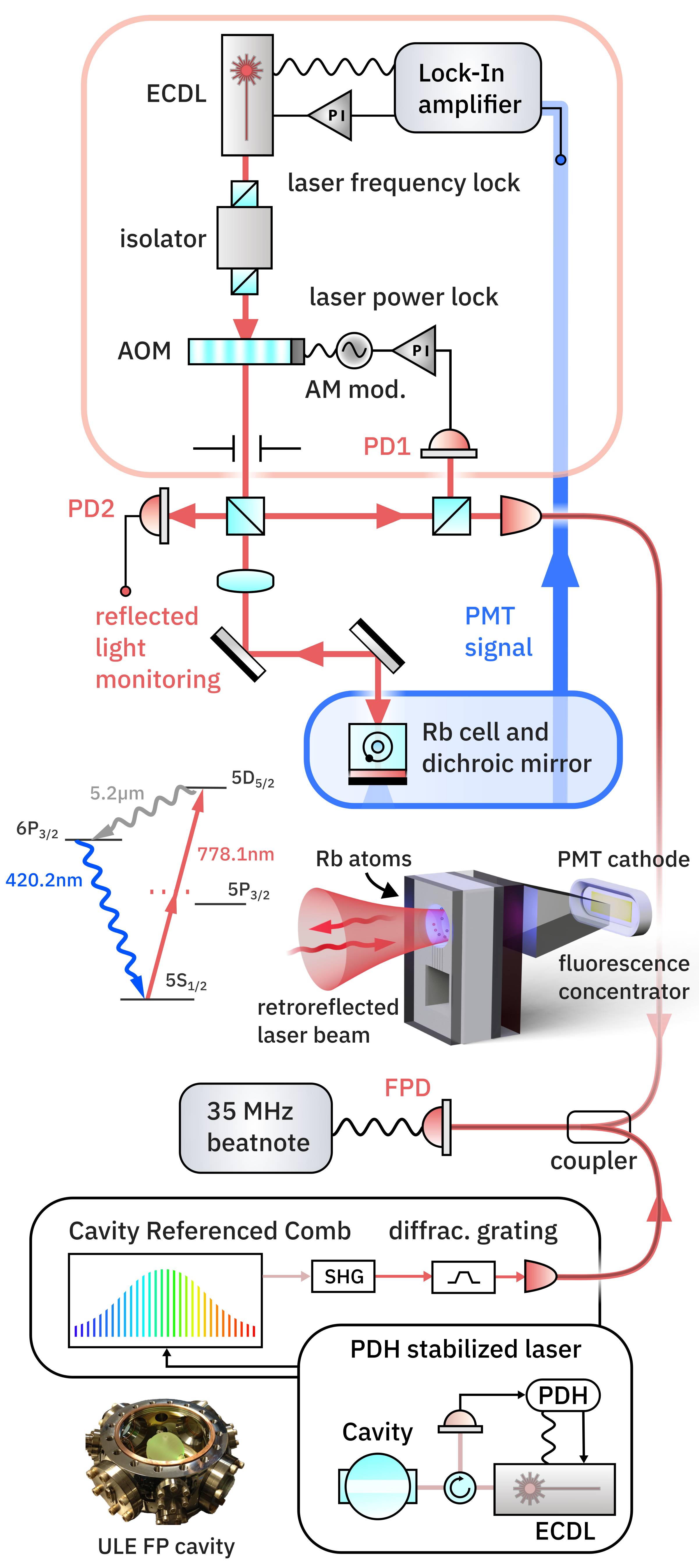}}
\caption{Experimental setup. An ECDL is frequency stabilized to the 778.1 nm two-photon absorption of the Rb atoms in a microfabricated vapor cell. After the excitation of the $5S_{1/2} \rightarrow 5D_{5/2}$ transition, atoms decay to the ground state via radiative decay, emitting 420 nm fluorescence which is then collected by a photomultiplier tube (PMT).
An acousto-optic modulator (AOM) is used to control and stabilize the laser power at the microcell input. A fraction of the locked laser light is injected into a fiber to be compared to the ultra-stable 778 nm signal extracted from a frequency-doubled frequency comb disciplined to a Ultra-Low-Expansion Fabry-Perot (ULE FP) cavity-stabilized laser. The final beatnote, obtained at about 35 MHz, is then analyzed using a frequency counter or phase noise analyzer. PD: photodiode. FPD: fast photodiode.}
\label{figure1}
\end{figure}

With a natural linewidth of about 330~kHz, the two-photon transition $5S_{1/2}\rightarrow 5D_{5/2}$ of the Rb atom at 778 nm constitutes an attractive option \cite{Nez:OC:1993}. In this scheme, shown as an inset in Fig. \ref{figure1}, Rb atoms are excited to the $5D_{5/2}$ state by means of the simultaneous absorption of two 778.1 nm photons and then decay back to the ground state via the $6P_{3/2}$ state, emitting in the process blue photons at 420.2~nm. The blue  fluorescence intensity is proportional to the 778.1 nm two photon absorption (TPA) and is then used to detect the sub-Doppler resonance peaks of rubidium. The presence of an intermediate $5P_{3/2}$ energy level close to the mid-point between the ground and excited levels increases the single-color two-photon absorption rate. This effect increases the signal-to-noise ratio (SNR) of the resonance signal, making this single-color approach a viable alternative to the complex, yet more efficient, 780~nm~-~776~nm two-color absorption scheme \cite{Gerginov:PRAp:2018, Perella:PRAp:2019}. The Rb two-photon absorption at 778~nm has been successfully used for the demonstration of vapor cell-based optical clocks with stability levels in the low 10$^{-13}$ range at 1 s and averaging down in the 10$^{-15}$ range \cite{Martin:PRAp:2018, Lemke:MDPI:2022}. A microcell optical clock operating on the Rb 778 nm TPA was demonstrated first at NIST \cite{Newman:O:2018}. A compact reference using a DBR laser was later reported with an Allan deviation of 2.8~$\times$~10$^{-12}$ at 1~s and approaching the 10$^{-13}$ level after a few 100 s \cite{Maurice:OE:2020}. Using a low noise ECDL and a more favorable line of the $^{85}$Rb isotope, this approach provided a stability of 1.8~$\times$~10$^{-13}$ at 1~s and approaching the 10$^{-14}$ level after a few 100~s \cite{Newman:OL:2021}.

In this work, we report on the development and short-term stability characterization of a microcell optical frequency reference based on the Rb two-photon absorption at 778~nm.  Measured against a 778~nm signal extracted from a cavity-stabilized laser, the optical standard offers an Allan deviation of 3.5~$\times$~10$^{-13}$ at 1~s, in good agreement with a laser phase noise of $+$~43~dBrad$^2$/Hz at 1~Hz Fourier  frequency, and averaging down to 2~$\times$~10$^{-14}$ at 200~s. The main limitation to the microcell reference short-term stability is currently the photon shot noise, explained by the limited number of blue photons collected at the detector. 
The contribution of the intermodulation effect \cite{Audoin:TIM:1991}, induced by the laser frequency noise, is estimated at the level of 1~$\times$~10$^{-13}$ at 1~s. 
These results confirm the potential of this spectroscopic approach for the development of high-stability microcell optical standards. Also, the analysis of the short-term noise budget indicates a considerable room for stability improvement of such microcell optical references.

\begin{figure*}[t]   
\centering
\includegraphics[width=0.9\linewidth]{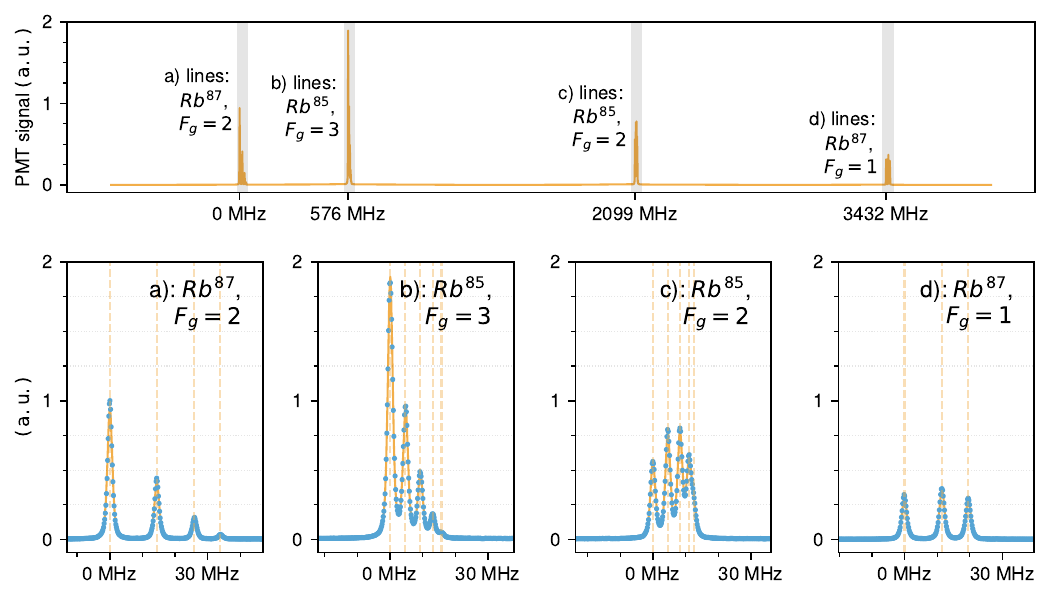}
\caption{Sub-Doppler spectroscopy of the $5S_{1/2}\rightarrow 5D_{5/2}$ transition at 778 nm of the Rb atoms in the microfabricated cell. The most intense peak is the $5S_{1/2} (F=3) \rightarrow 5D_{5/2} (F=5)$ transition of \rbisob{} atom (peak on the left of sub-plot (b)). However, its proximity with neighboring lines made its use for the laser lock not robust enough. In this work, we used the $5S_{1/2} (F=2) \rightarrow 5D_{5/2} (F=4)$ transition of \rbisoa{} atom (leftmost peak in sub-plot (a)). For each sub-plot on the bottom, which zooms on the hyperfine structure of \rbisoa{} and \rbisob{} lines, the first peak on the left was arbitrarily fixed to a null laser frequency detuning.}
\label{fig:full}
\end{figure*}

\section{Experimental setup}
Figure \ref{figure1} shows a schematic of the experimental setup. The microcell optical reference consists of an ECDL, stabilized to the $5S_{1/2} (F_g = 2) \rightarrow 5D_{5/2} (F_e =4)$ two-step transition of rubidium at 778.1~nm.
A Faraday optical isolation stage in front of the laser reduces optical feedback. An acousto-optic modulator (AOM) is then used for the tuning and stabilization of the total laser power \cite{Tricot:RSI:2018}, independently of laser bias current. For this purpose, a fraction of the laser beam is directed through a beam-splitter cube onto a photodiode (PD1). The output signal of this photodiode is compared to a voltage reference in order to produce an error signal. This signal is then fed to a proportional-integral (PI) controller that adjusts the amplitude of the RF signal used to drive the AOM. Most of the laser power at the output of the AOM (12~mW) is sent to a physics package  that houses a microfabricated Rb vapor cell, complete with temperature regulation and mu-metal magnetic shielding. We name the total laser power at the cell input $P_l$. A second photodiode (PD2) measures the fluctuations of the reflected laser beam. 
\begin{figure}[t]   
{\centering\includegraphics[width=0.9 \linewidth]{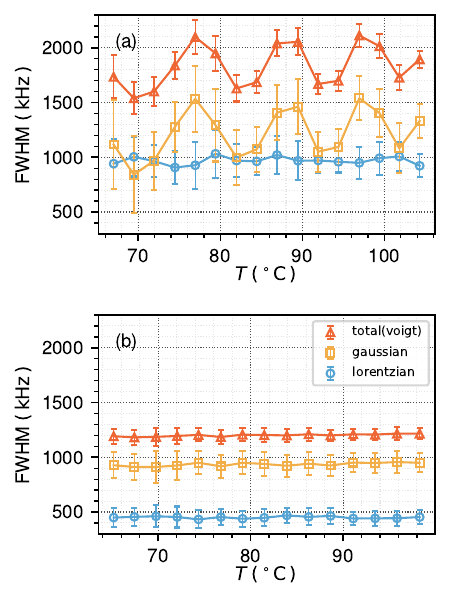}}
\caption{
Linewidth of the $F_g=2$, $F_e = 4$ line of \rbisoa{} isotope versus the cell temperature, for the microfabricated cell in (a) and for a reference glass-blown evacuated Rb vapor cell in (b). The total linewidth, as well as Lorenztian and Gaussian linewidth contributions, have been estimated by fitting the lineshape of the resonance.
Error bars, representing $2 \sigma$ uncertainty, are extracted from the statistical distribution of 400 lineshape fits.
}
\label{fig:massfit}
\end{figure}

The MEMS cell fabrication followed the processes described in \cite{Hasegawa:SA:2011, Vicarini:SA:2018}, yet no buffer gas was introduced before the cell sealing. The cell consists of two neighboring cavities, structured in silicon using deep-reactive ion etching (DRIE), and sandwiched between two anodically-bonded 500~$\mu$m-thick borosilicate glass windows. A pre-embedded Rb pill-dispenser alloy, laser-activated after final sealing of the cell, is used to fill the cell with Rb vapor (natural isotopic abundance). The cavity in which atom-light interaction takes place has a diameter of 2 mm and a length of 1.5~mm. The cell can be temperature-stabilized at set-points ranging from 70 to 120$^{\circ}$C. A plano-convex lens ($f\,=\,$100~mm), placed in front of the cell, is used to focus the laser beam so that the waist of the laser beam ($w_0$~$\sim$~100~$\mu$m) is located at the surface of a dichroic mirror located on the backside of the cell. This lens-mirror arrangement ensures a high light intensity in the cell and the overlapping of counter-propagating laser beams for Doppler broadening suppression. The mirror reflects the near-infrared (NIR) laser light but allows the blue 420~nm fluorescence light to pass through. A fluorescence concentrator, that consists of a lens doublet and a band-pass filter at 420~nm, is finally used to collect the fluorescence light onto a photomultiplier (PMT) while rejecting the detection of spurious NIR laser light. The resonance signal detected by the PMT is amplified with a high-gain transimpedance amplifier and fed to a lock-in amplifier to generate a dispersive zero-crossing error signal. 
This error signal is then used to generate a PID correction signal applied to the ECDL bias current, locking the laser to the atomic resonance frequency.

In parallel, a fraction of the laser output is coupled into an optical fiber and sent to an adjacent laboratory. There, it is used to detect a beatnote signal against a reference 778.1~nm signal delivered by a frequency-doubled, femtosecond-laser optical frequency comb (OFC) disciplined to an ultra low expansion (ULE) Fabry-Perot cavity stabilized laser.
Long term drift due to the cavity is compensated thanks to the long term stability of the 100 MHz signal from a hydrogen maser. This drift control is performed with a phase-locked loop (PLL) that makes use of the OFC to realize the frequency division from the optical domain to the RF frequency range. With drift compensation, the cavity-based optical reference exhibits a short-term frequency stability of 2~$\times$~10$^{-15}$ at 1~s and below the 10$^{-14}$ level at 10$^4$~s \cite{Didier:AO:2015}. These performances are sufficient to ensure that frequency fluctuations of the beatnote are due to the microcell optical reference under test.

\section{Experimental results}
\subsection{Features of the atomic resonance}
Figure \ref{fig:full} shows the full spectrum of the $5S_{1/2}\rightarrow 5D_{5/2}$ transition at 778 nm of the Rb atoms in the microcell. Both \rbisoa{} and \rbisob{} are present, with natural isotopic ratio. When hyperfine splitting is taken into account, absorption lines can be classified by the isotope and by the total angular momentum of the initial and final states of the transition, $F_g$ and $F_e$. Among all these resonances, the $F_g=3 \rightarrow F_e = 5$ line of \rbisob{} isotope (see Fig. \ref{fig:full}(b)) is the most intense and thus an optimal choice for the development of the optical frequency reference \cite{Newman:OL:2021}. 
However, the presence of broadening due to contaminants in the cell and the close proximity of neighboring hyperfine levels render the frequency lock sensitive to line pulling and frequency hopping, compromising its robustness against random disturbances.
Instead, the resonance with the second highest intensity, the $F_g=2$, $F_e = 4$ line of the \rbisoa{} isotope (peak on the left of Fig. \ref{fig:full}(a)), with a reduced overlap between resonances, is used. 

\begin{figure}[ht]   
{\centering\includegraphics[width=0.9\linewidth]{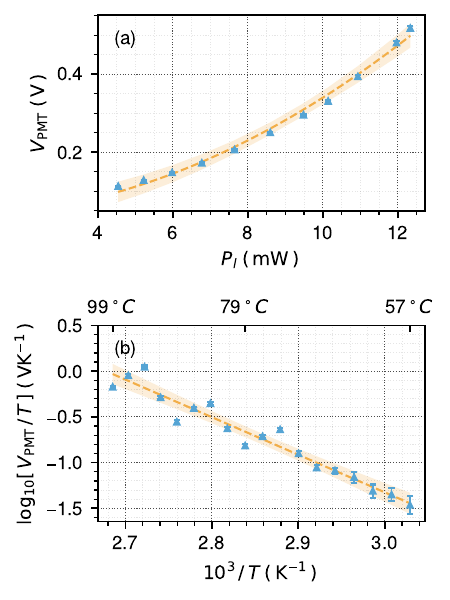}}
\caption{
Amplitude of the two-photon resonance versus the laser power $P_l$ at the cell input (a) and the inverse of the cell temperature (b). For (a), the cell temperature was set to $110 {}^\circ C$. A second order polynomial is used to fit experimental data. For (b), the laser power $P_l$ was fixed to 12.3~mW. \eqref{eq:antoine} is used to fit the data    to the vapor pressure. The parameter $A$ has been adapted to group together all offsets present in the logarithmic scale plot for a simplified interpretation. In contrast, the parameter $B$ of \eqref{eq:antoine} is used according to its original definition. The error bars and confidence intervals are drawn for $2 \sigma$ uncertainty. The discrepancy observed at high temperatures can be explained by thermal effects in the MEMS cell holder (see Fig. \ref{fig:massfit}).
}
\label{fig:ptcell}
\end{figure}
\begin{figure}[t]   
{\centering\includegraphics[width=\linewidth]{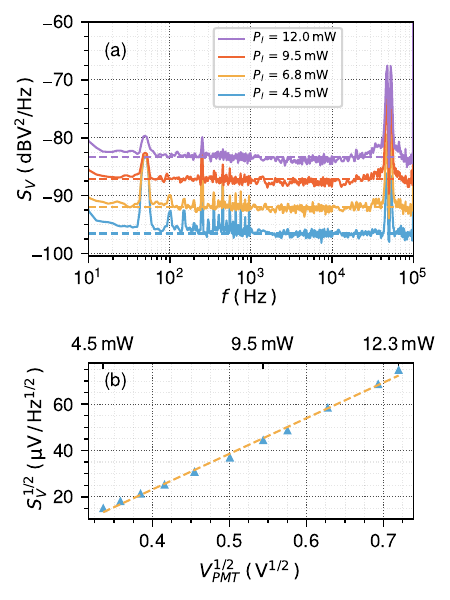}}
\caption{(a) Total detection noise density spectra measured with a FFT analyzer at the output of the PMT for different laser powers at the cell input. Dashed horizontal lines are used to estimate the white noise floor. (b) Detection white noise density at the PMT output versus the square-root of the PMT output signal amplitude.
The cell temperature is held constant at 110 ${}^\circ$C.
}
\label{fig:shot_noise}
\end{figure}
\begin{figure}[t]   
{\centering\includegraphics[width=\linewidth]{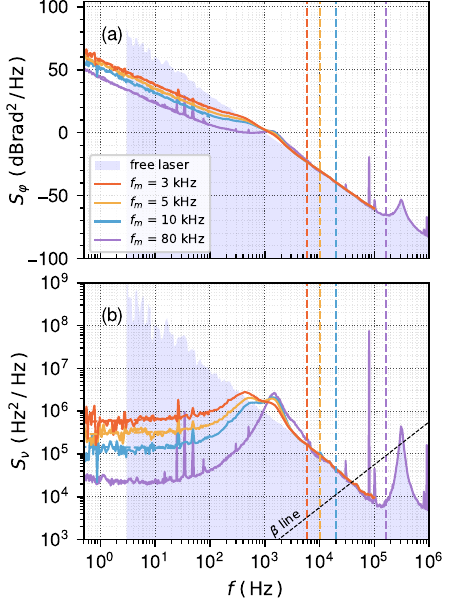}}
\caption{(a) Absolute phase noise of the beatnote ($\sim$ 35 MHz) between the microcell-stabilized laser and the cavity-stabilized 778 nm reference signal, obtained for different laser modulation frequencies $f_m$. 
Vertical dashed lines are used to indicate $S_\varphi(2 f_m)$.
The shaded area shows the phase noise of the ECDL in the free-running regime (not locked to the microcell). (b) Frequency noise spectra, for the same conditions as in (a).
The bump at $\sim$ 200 kHz is due to the ULE cavity servo loop.
The shaded area shows the free-running frequency noise of the ECDL used in the TPA setup. The $\beta$ separation line \cite{DiDomenico:AO:2010} is included for reference.  Spectra were obtained using a phase noise analyzer (RS FSWP). 
}
\label{fig:pn}
\end{figure}
Fitting the lineshape of the peaks in Fig. \ref{fig:full} reveals a Voigt line profile, which arises from a convolution of Lorenztian and Gaussian broadening components. The Lorentzian linewidth results from homogeneous broadening, due to the natural linewidth of the transition, self-broadening, and collisional broadening induced by the presence of impurities in the cell. The Gaussian linewidth results from inhomogenous broadening, mainly due to residual Doppler broadening caused by an imperfect overlap between the two counter-propagating beams of the Doppler-free scheme. Figure \ref{fig:massfit}(a) shows the total linewidth of the atomic resonance, as well as the Gaussian and Lorentzian widths, versus the cell temperature. For comparison, data obtained in an evacuated glass-blown Rb vapor cell, of diameter 25~mm and length 70~mm, are reported in Fig. \ref{fig:massfit}(b). In the MEMS cell, the total linewidth is measured to be in the 1.5 - 2.1 MHz range. 
The Gaussian linewidth swings retrace across both increasing and decreasing temperature sweeps and are likely a result of incomplete cancellation of the Doppler effect, caused  by misalignment between the counter-propagating laser beams.
Notably, these oscillations are observed in the MEMS package but are absent in the glass-blown cell, suggesting that they are due to thermal expansion in the MEMS cell holder.
The Lorentzian width in the MEMS cell is slightly below 1~MHz. This value, higher than the transition natural linewidth ($\sim$~330 kHz), is mainly attributed to collisional broadening from gas impurities.
In the glass-blown cell, the Lorentzian width is equal to 450~kHz. The main broadening ($\sim$~100 kHz) in this case may be explained by the permeation of atmospheric helium ($\sim$~4~mTorr) \cite{Zameroski:JP:2014}. For both types of cells, the Gaussian linewidth contribution is slightly increased with the rise of the cell temperature. Note that we did not observe any significant variation of the resonance linewidth with the laser intensity.

The amplitude of the atomic signal is correlated with both the intensity of the laser beam used to probe the transition and the rubidium vapor density.
Two-photon absorption, being a second order nonlinear optical process, is proportional to the square of the incident laser intensity, provided that the atomic transition is not saturated. Additionally, the detection of 420 nm fluorescence is influenced by the branching ratio of the decay from the 5D$_{5/2}$ excited state.
This ratio must remain constant to ensure that the fluorescence signal is proportional to the two photon absorption. This behaviour is well observed on Fig. \ref{fig:ptcell}(a).

On the other hand, the density of the rubidium vapor increases exponentially with temperature, as described by the simplified Antoine equation \cite{Thomson:CR:1946}
\begin{equation}
\label{eq:antoine}
\begin{aligned}
    \log_{10}{P_v} &= A - B /T \, ,
\end{aligned}
\end{equation}
where $P_v$ is the vapor pressure, which can be related to the vapor density by means of the ideal gas law, $T$ is the cell temperature and $A$ and $B$ are empirical parameters. Figure \ref{fig:ptcell}(b) shows the intensity of the atomic resonance, assumed to be proportional to the vapor density, and the fit to the Antoine equation. The value of $B$ obtained from the fit of experimental data is  $B_\mathrm{exp} = -4100 \pm 260$ K , matching the reference value found in the literature ($ B_\mathrm{ref}~=~-4040~\pm~200$~K) \cite{Alcock:CMQ:1984}.
In light of these results, we have chosen to operate at $T=$~110~$^{\circ}$C and $P_l=$~12.3~mW (the maximum available laser power) to optimize the signal to noise ratio.

\subsection{Main noise sources}
\subsubsection{Photon shot noise}
Once we have assessed the influence of key experimental parameters on the resonance signal and linewidth, we can evaluate the main noise sources that affect the atomic resonance.
Photon shot noise, originating from the quantization of light, is usually recognized as an important limiting factor to the short-term stability of optical standards based on TPA, due to the low absorption efficiency of this process.
This type of noise exhibits white noise power spectral density (PSD) and a variance proportional to the light intensity. If this noise source dominates, the signal-to-noise ratio (SNR) is then directly proportional to the square root of the intensity of the signal.
Furthermore, setting aside other sources of noise, the photodetector current noise PSD at the PMT cathode will be given by $S_i(f) = 2qI$ , where $q$ is the charge of the electron and $I$ is the mean value of the photocurrent.

\begin{figure}[t]   
{\centering\includegraphics[width=\linewidth]{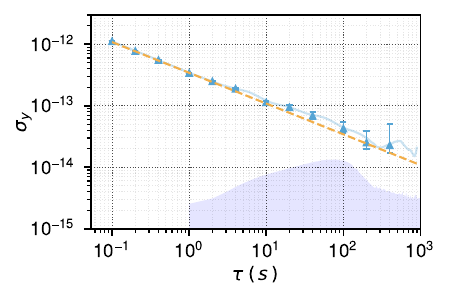}}
\caption{
Short-term fractional frequency stability of the microcell-stabilized laser, measured by beating its output signal against the reference cavity-stabilized 778 nm signal. The cell temperature is held constant at 110 ${}^\circ$C, and the laser power $P_l$ is equal to 12.3 mW. The shaded area shows the stability level of the cavity-stabilized reference. The microcell optical reference demonstrates a fractional Allan deviation of 3.5 $\times$ 10$^{-13}$ $\tau^{-1/2}$ until 200 s (dashed line), averaging down to the low 10$^{-14}$ range. A continuous line with the overlapping Allan deviation is also included. The value of $\sigma_y(1s)$ is consistent with the value reported in Fig. \ref{fig:adev_snr}.
}
\label{fig:adev}
\end{figure}

\begin{figure}[t]   
{\centering\includegraphics[width=\linewidth]{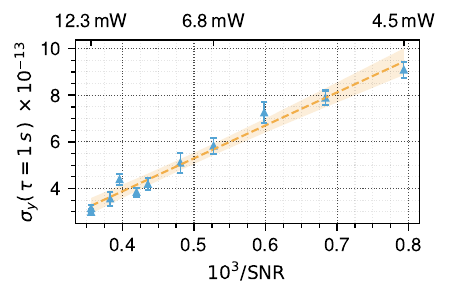}}
\caption{
Short-term stability at 1 s of the microcell-stabilized reference versus 1/SNR.
The cell temperature is held constant at 110 ${}^\circ$C, the amplitude SNR values (evaluated in a 1 Hz bandwidth) are the same as those in Fig. \ref{fig:shot_noise}, and are obtained by changing the laser power, as shown in the top x-axis.
Experimental data are fitted to \eqref{eq:stab}, showing a good agreement with the SNR limited noise model.
The error bars and confidence intervals are drawn to represent the $2 \sigma$ uncertainty.
}
\label{fig:adev_snr}
\end{figure}

Figure \ref{fig:shot_noise}(a) shows detection noise spectra measured at the output of the PMT transimpedance amplifier, for different values of the laser power at the cell input. We clearly observe a shift of the noise floor (white noise) following each increment in the laser power, together with a change in the amplitude of the frequency lock servo sidebands around 80 kHz. 
Derived from such spectra, Fig. \ref{fig:shot_noise}(b) compares the detection noise floor $S_v$ to the shot noise model (noise power $N \propto S^{1/2}$), revealing a shot noise limited signal.

\subsubsection{Intermodulation effect}
Another significant contribution to the short-term stability of passive continuous-wave optical frequency references is the intermodulation effect, induced by the frequency noise of the interrogating laser \cite{Audoin:TIM:1991}.
This short-term instability limit is given by
\eqref{eq:intermod},
with $S_{\nu}(2f_m)$ the power spectral density of frequency fluctuations of the laser at twice the modulation frequency $f_m$, and $\nu_0$ the clock transition frequency ($\nu_0 = $ 3.85$\,\times\,$10${}^{14}$~Hz in this case)
\begin{equation}
\label{eq:intermod}
\sigma_y (\tau) \simeq \frac{1}{2} \sqrt{\frac{S_{\nu}(2f_m)}{\nu_0^2}} \frac{1}{\sqrt{\tau}} \, .
\end{equation}

Figure \ref{fig:pn} shows the phase noise (a) and frequency noise (b) spectrum of the ECDL used to probe the TPA resonance, in free-running and locked regimes. In the free-running case, the phase noise of the ECDL is about $-$~65~dBrad${}^2$/Hz 
at $f$~=~100~kHz, and $\sim$~90~dBrad$^2$/Hz (10$^{9}$~rad${}^2$/Hz) at $f$ = 1 Hz. In the locked case, spectra were recorded for different values of the laser modulation frequency, $f_m$. Best results, obtained here with $f_m$~$\simeq$~80~kHz, yield a phase noise of about $+$~43~dBrad${}^2$/Hz at $f$~=~1~Hz. Looking at Fig. \ref{fig:pn}(b), the frequency noise of the ECDL in the free-running case is about 6000  Hz${}^2$/Hz at $f$~=~10${}^5$ Hz. With $f_m$~=~10${}^5$~Hz, the instability limit due to the intermodulation effect is then estimated at the level of 1.0~$\times$~10${}^{-13}$ at 1~s. We also note on Fig. \ref{fig:pn}(b) that the laser frequency servo bandwidth is around 2 kHz. In addition to the main result, $f_m$~=~80~kHz, three traces are shown for $f_m=$3, 5, 10 kHz (the $f_m$ spikes have been removed for clarity). The effect of the intermodulation noise on the locked frequency noise is clear although it is not the limiting factor. We note that for these three measurements, the servo bandwidth was reduced by adjusting the PID controller.

\subsection{Short-term frequency stability}
We have measured the short-term fractional frequency stability of the microcell-stabilized ECDL. Corresponding results are reported in Fig. \ref{fig:adev}. The Allan deviation of the microcell reference is $\sigma_y(\tau)$~=~3.5~$\times$~10${}^{-13}$~$\tau^{-1/2}$ until 200~s. This stability is approximately two times worse than the one reported in \cite{Newman:OL:2021}, in which the twice more intense $F_g=3$, $F_e = 5$ line of \rbisob{} isotope was used, and where the minimum laser frequency noise was slightly lower (5~$\times$~10$^3$~Hz$^2$/Hz). Hence, these results confirm the potential of this spectroscopic approach for the advancement of ultra-stable microcell-stabilized optical frequency references.

The SNR limited short-term stability of the optical microcell reference can be approximated by \eqref{eq:stab}, with $\eta$ a semi-empirical term \cite{VanierAudoin} that represents the linearized response of the feedback scheme, $\Delta \nu$ the resonance linewidth ($\sim$ 1.5 MHz), $S$ the resonance amplitude and  $S_v$ the detection noise amplitude density

\begin{equation}
\sigma_y(\tau) \simeq \eta \frac{\Delta \nu}{\nu_0} \frac{S_v}{S}  \frac{1}{\sqrt{ 2 \tau}} \, .
\label{eq:stab}
\end{equation}
Simultaneously with the SNR measurement (with $SNR = S/N$ and total noise power $N$ evaluated for a 1 Hz bandwidth) illustrated in Fig. \ref{fig:shot_noise}, we used a frequency counter to measure the beatnote frequency between the microcell reference and the comb. The agreement between the measurements and \eqref{eq:stab}, illustrated by the linear fit of the stability at 1~s as a function of 1/SNR in Fig. \ref{fig:adev_snr}, confirms that the short-term frequency stability of the microcell optical standard is currently limited by the SNR, at all laser power values.
In addition, the value of the slope of the linear fit, equal to $m=1.4 \pm 0.1 \times 10^{-9}$, can be used to estimate $\eta = 0.43 \pm  0.04$.

\subsection{Power light-shift}
\begin{figure}[t]   
{\centering\includegraphics[width=\linewidth]{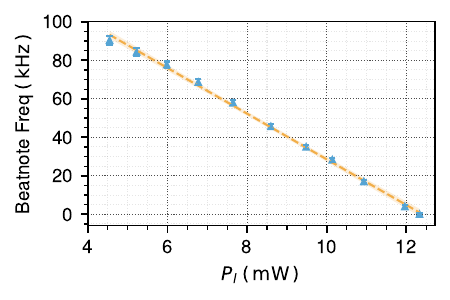}}
\caption{
Dependence of the microcell reference frequency to the laser power at the cell input, $P_l$. Experimental data are fitted by a linear function, with a slope of $-$ 11.9 $\pm$ 0.3 kHz/mW.
The cell temperature is held constant at 110 ${}^\circ$C.
Error bars and confidence intervals are drawn to represent the $2 \sigma$ uncertainty.
}
\label{fig:lshift}
\end{figure}

Beyond 100 seconds of integration time, a degradation of the microcell reference stability is observed.
Among potential contributions, light-shift effects have been identified in several studies \cite{Maurice:OE:2020, Martin:PRAp:2018, Lemke:MDPI:2022} as an important contribution to the mid-term and long-term stability of optical cell clocks based on TPA. As a last investigation in this study, we have then measured the dependence of the microcell reference frequency to variations of the laser power at the cell input. Results are illustrated in Fig. \ref{fig:lshift}.

Experimental data are well-fitted by a linear sensitivity coefficient of $-$ 11.9~$\pm$~0.3~kHz/mW, i.e. $-$~3.1$\times$10$^{-11}$/mW in fractional value. This coefficient, about twice larger than the one measured in \cite{Newman:OL:2021} (explained by a different laser waist size),
indicates that fluctuations of the laser power should be reduced at a level lower than 3.2$\times$~10$^{-5}$ to reach a stability level of 10$^{-15}$. 
Light-shift mitigation techniques might be also implemented to reduce the power light-shift coefficient \cite{Yudin:PRAp:2020, MAH:PRAp:2020, Kang:OE:2024}. Variations on the pointing direction of the laser beam could also induce a random, time-dependent light shift proportional to the intensity of the laser light. The reduction of collisional shifts, sensitive to cell temperature fluctuations, will also require careful attention. These studies remain out of the scope of the paper and will be performed in a future work.

\section{Conclusions}
We have reported the development and short-term stability characterization of an external-cavity diode laser stabilized on a microfabricated vapor cell using spectroscopy of the Rb 778 nm two-step transition. The impact of some key experimental parameters (laser power, cell temperature) on the sub-Doppler resonance features (signal and linewidth) was reported. The microcell-stabilized laser demonstrates an Allan deviation of 3.5~$\times$~10${}^{-13}$~$\tau^{-1/2}$, reaching 2~$\times$~10${}^{-14}$ at 200~s. The laser short-term instability is in good agreement with the resonance signal-to-noise ratio and is currently limited by the photon shot noise. It could be improved with increased laser power or cell temperature.
The contribution of the intermodulation effect, induced by the free-running laser frequency noise (6000~Hz${}^2$/Hz at $f$~=~1.6~$\times$~10${}^5$ Hz), is evaluated at the level of 1~$\times$~10$^{-13}$ at 1~s.
Various paths of optimization, such as the development of microcells with enhanced purity \cite{Boudot:SR:2020, Martinez:NC:2023} for the detection of narrower resonances, the optimization of the fluorescence collection and the use of the more intense $5S_{1/2} (F=3) \rightarrow 5D_{5/2} (F=5)$ transition of $^{85}$Rb atom \cite{Newman:OL:2021} for atomic signal enhancement, or the use of lasers with lower frequency noise \cite{Guo:SA:2022, Clementi:LSA:2023}, might be explored in the future for improvement of the microcell reference short-term stability. Studies will be also engaged to mitigate main frequency shifts and improve the mid-term and long-term stability of the optical standard.

\section*{Funding}
This work was partly funded by Centre National d'Etudes Spatiales (CNES) in the frame of the OSCAR project (Grant 200837/00). This project was partly funded by the European Union. Views and opinions expressed are however those of the author(s) only and do not necessarily reflect those of the European Union. Neither the European Union nor the granting authority can be held responsible for them. This project has received funding from the European Defence Fund (EDF) under grant agreement EDF-2021-DIS-RDIS-ADEQUADE. This work was also supported by the Agence Nationale de la Recherche (ANR) in the frame of the LabeX FIRST-TF (Grant ANR 10-LABX-48-01), the EquipX Oscillator-IMP (Grant ANR 11-EQPX-0033) and the EIPHI Graduate school (Grant ANR-17-EURE-0002) through the REMICS project.  This work was partly supported by the french RENATECH network and its FEMTO-ST technological facility (MIMENTO).

\section*{Acknowledgments}
The authors would like to thank V. Giordano and M. Abdel Hafiz for careful reading of the manuscript and M. Hauden for his valuable input.

\section*{Disclosures}
The authors declare no conflicts of interest.

\section*{Data availability statement}
The data that support the findings of this study are available
from the corresponding author upon reasonable request.

\end{document}